\documentclass[aps,pra,twocolumn,showpacs]{revtex4}

\usepackage{graphicx,braket}
\newcommand{\up}{\uparrow}
\newcommand{\down}{\downarrow}

\begin{document}

\title{Decoherence of a two-qubit system with a variable bath coupling operator}
\author{M.J.\ Storcz\footnote{Email: storcz@theorie.physik.uni-muenchen.de}}
\author{F.\ Hellmann}
\author{C.\ Hrelescu}
\author{F.K.\ Wilhelm}
\affiliation{Physics Department, Arnold Sommerfeld Center for Theoretical
Physics, and Center for NanoScience, Ludwig-Maximilians-Universit\"at
M\"unchen, 80333 M\"unchen, Germany}

\begin {abstract}
We examine the decoherence of an asymmetric two-qubit system that is coupled via a tunable interaction term
to a common bath or two individual baths of harmonic oscillators. The dissipative dynamics are evaluated
using the Bloch-Redfield formalism. It is shown that the behaviour of the decoherence effects is affected mostly
by different symmetries between the qubit operator which is coupled to the environment and temperature, whereas
the differences between the two bath configurations are very small.
Moreover, it is elaborated that small imperfections of the qubit parameters do not lead to a drastic enhancement
of the decoherence rates.
\end {abstract}


\maketitle

\section{Introduction}
Quantum computation provides a substantial speedup for several important computational tasks \cite{shorsalgorithm,grover,deutsch,
divincenzo1}. A general quantum bit (qubit) consists of a two level quantum system with a controllable Hamiltonian
of sufficient generality to implement a universal set of quantum logic gates \cite{nielsenchuang}. From such
a set, an arbitrary quantum algorithm can be implemented to any desired accuracy limited only by decoherence.
A universal two qubit system requires just single-qubit rotations and one additional entangling two qubit gate.
One important example is the controlled-NOT ({\sc cnot}) gate which switches the state of the second qubit depending
on the state of the first qubit and is given in spinor representation of $\hat \sigma^{(1)} \otimes \hat \sigma^{(2)}$ by
\begin {equation}
U_{\rm CNOT} =  \left( \begin {array}{cccc}
1 & 0 & 0 & 0 \\
0 & 1 & 0 & 0 \\
0 & 0 & 0 & 1 \\
0 & 0 & 1 & 0 \\
\end {array} \right)\textrm{,}
\end {equation}
up to a global phase.

Superconducting Josephson charge and persistent current (flux) qubits have been shown to possess the necessary properties \cite{divincenzo1}
to act as a quantum bit. They have been manipulated coherently and coherence times in the $\mu s$ range have been demonstrated
\cite{orlando,thesiscaspar,pashkin,nakamura,quantronium,fluxqubitrabi,cnotscb} with a corresponding quality factor
of quantum coherence of up to $Q_\varphi \approx 10^4$ \cite{quantronium}.
In a two qubit system, where the coupling was achieved using a shared Josephson junction, coherent Rabi oscillations
between states of a coupled qubit system were observed \cite{paauw,alex} and in a two charge qubit system a conditional
gate operation was performed \cite{cnotscb}.
All of these experiments suffer from material imperfections which lead to non-ideal time evolutions of the quantum states
due to a parameter spread in the characteristic energies of the system Hamiltonian.
On the other hand, for high symmetry of the qubit parameters, the qubit
coherence can be strongly protected. This extends from the protection 
of the singlet in a symmetric qubit setup \cite{pra} to the general concept
of decoherence-free subspaces (DFS) \cite{kempe,preprint}. General considerations
on the stability of such DFS can be found in Ref.\ \onlinecite{Bacon99}.
In this paper, the dependence of the decoherence rates and gate quality factors 
on the parameter spread of the qubits will be elaborated theoretically. In 
perspective, this is of crucial importance for connecting the experimental
status and prospects to these central concepts in quantum information science: 
which degree of parameter uniformity do experiments have to achieve for 
symmetry-based protection schemes to work --- do these schemes have to be 
extended in order to accomodate experimental restrictions?

Moreover, variable bath couplings to the decohering environment have also been identified, {\em e.g.}, in Ref.\,\cite{zorin1}.
The decoherence properties of different qubit operators that are coupled to the environmental bath will be
investigated in this work.

The Hamiltonian of a typical pseudo-spin system can be expressed in terms of the Pauli matricies as
\begin{equation}
\mathbf{H}_{\rm q} = - \frac{1}{2}(\epsilon~ \hat \sigma_z +\eta~ \hat \sigma_x)\textrm{.}
\end{equation}
In a two qubit system, an additional interaction term is required to
implement the universal two qubit gate. In superconducting implementations
\cite{orlando,pashkin,nakamura,stonybrook,quantronium,fluxqubitrabi,makhlin1} this
coupling term is typically proportional to $\hat \sigma_z^{(1)}~ \hat \sigma_z^{(2)}$. Here, the
superscripts are the qubit indices. In particular,
inductively coupled flux qubits \cite{ftq,pra} and capacitively
coupled charge qubits \cite{pya} are coupled this way. Thus, the two
qubit Hamiltonian is
\begin{equation}
\mathbf{H}_{\rm 2q} = -\frac{1}{2}\sum_{i=1}^2 \limits \left(\epsilon^{(i)}~ \hat \sigma_z^{(i)}
+\eta^{(i)}~ \hat \sigma_x^{(i)}\right)- \frac{\gamma}{2}~ \hat \sigma_z^{(1)}~ \hat \sigma_z^{(2)}\textrm{.}
\end{equation}
In the singlet/triplet basis, $(1,0,0,0)^T = \ket{\up\up}$,
$(0,1,0,0)^T = (\ket{\up\down} + \ket{\down\up})/\sqrt{2}$,
$(0,0,1,0)^T = \ket{\down \down }$,
$(0,0,0,1)^T = (\ket{\up \down } - \ket{\down \up })/\sqrt{2}$
that exhibits the symmetry properties of the coupling most clearly, this Hamiltonian
takes the following explicit matrix form
\begin {equation}
\mathbf {H}_{\rm 2q} = -\frac{1}{2} \left( \begin {array}{cccc}
\epsilon& \eta & \gamma & -\Delta\eta \\
\eta & -\gamma & \eta & -\Delta\epsilon \\
\gamma & \eta & -\epsilon & \Delta\eta \\
-\Delta\eta & -\Delta\epsilon & \Delta\eta & \gamma \\
\end {array} \right)\textrm{,}
\end {equation}
with $\epsilon = \epsilon^{(1)}+\epsilon^{(2)}$, $\Delta\epsilon = \epsilon^{(1)}-\epsilon^{(2)}$
and $\eta = (\eta^{(1)}+\eta^{(2)})/\sqrt{2}$, $\Delta\eta = (\eta^{(1)}-\eta^{(2)})/\sqrt{2}$.
Using this Hamiltonian the {\sc cnot} gate can be implemented through a sequence
of elementary quantum gates \cite{thorwart,pra}
\begin {eqnarray} \label{cnotgateseq}
U_{\rm CNOT} & = &
U_{\rm H}^{(2)}
\exp \left(-i \frac{ \pi }{4} \hat \sigma_z^{(1)}\right)
\exp \left(-i \frac{ \pi }{4} \hat \sigma_z^{(2)}\right) {} \nonumber \\
& &{} \times
\exp \left(-i \frac{ \pi }{4} \hat \sigma_z^{(1)}\sigma_z^{(2)}\right)
\exp \left(-i \frac{ \pi }{2} \hat \sigma_z^{(1)}\right)
U_{\rm H}^{(2)}
\textrm{, }\quad
\end {eqnarray}
where $U_{\rm H}^{(2)}$ denotes the Hadamard gate operation performed on the second qubit.
It involves one two-qubit operation at step three only. For our numerical calculations we applied
the characteristic energies that were used in \cite{pra} as a viable example for superconducting
solid-state flux or charge qubits. Following this approach also gate sequences optimized with
respect to decoherence have been studied \cite{optdec}.

Disregarding the Hadamard gates, the gate operation Eqn. (\ref{cnotgateseq}) forms a controlled-phase ({\sc cphase}) gate
\begin {equation}
U_{\rm CPHASE} =  \left( \begin {array}{cccc}
1 & 0 & 0 & 0 \\
0 & 1 & 0 & 0 \\
0 & 0 & 1 & 0 \\
0 & 0 & 0 & e^{i\phi} \\
\end {array} \right)\textrm{,}
\end {equation}
with $\phi = \pi$.
\\
In experimental realizations of this model, additional effects always impair the capability of the system
to operate as a qubit. In condensed matter implementations, the most pronounced is the coupling
to environmental degrees of freedom. This leads to relaxation, {\em i.e.}, classical thermalization of the states as well
as, on a much shorter timescale, to dephasing. Decoherence causes the system to act like a classical ensemble
eliminating all potential computational benefits of quantum algorithms. For a wide range ({\em e.g.} \cite {governale, pra, pya})
of solid state implementations the dominant decoherence effects caused by coupling to
linear environments such as electric circuits obey Gaussian statistics and can be effectively modeled with a bath of
harmonic oscillators. It is assumed here that there is only one decoherence source in the
dominating order of magnitude in the coupling parameter and possible weaker
noise sources are ignored.
To model this source each qubit is either coupled to an individual
or to a common bath of harmonic oscillators. The system Hamiltonian then takes the form
\begin{equation}
\mathbf{H}_{\rm 2qb}^{\rm 2B} =  \mathbf{H}_{2q} + \frac{1}{2}\left( \hat \sigma_s^{(1)}~ \hat X^{(1)} +
\hat \sigma_s^{(2)}~ \hat X^{(2)}\right) + \mathbf{H}_B^{(1)} + \mathbf{H}_B^{(2)}
\end{equation}
or
\begin{equation}
\mathbf{H}_{\rm 2qb}^{\rm 1B} = \mathbf{H}_{2q} + \frac{1}{2}~\left(\hat \sigma_s^{(1)} + \hat \sigma_s^{(2)}\right)~ \hat X + \mathbf{H}_B\textrm{,}
\end{equation}
where $\hat \sigma_s$ is the spin-representation of the qubit operator talking to the environment that depends on the specific implementation
of the qubit. For the special case of superconducting flux qubits which only experience flux noise and superconducting charge qubits which are only subject
to charge noise this would correspond to $\hat \sigma_s = \hat \sigma_z$. Here, $\hat X$ is the collective coordinate of
the harmonic oscillator bath and the superscript distinguishes between the single bath and the two bath case.
The general form is
\begin{equation}
\hat \sigma_s = (\vec{c}\cdot \vec{\sigma}) = \sqrt{2}(c_x~\hat \sigma_x + c_y~\hat \sigma_y) + c_z\hat \sigma_z
\textrm{,}
\end{equation}
where the factor $\sqrt{2}$ in front of $c_x$ and $c_y$ was chosen for convenience in the singlet/triplet
basis in which the qubit-bath interaction becomes
\begin {equation}
\mathbf {H_{int}} = \frac{1}{2} \left( \begin {array}{cccc}
c_z\, X & c_-\, X & 0 & -c_-\, \Delta X \\
c_+\, X  & 0 & c_-\, X  & c_z\, \Delta X \\
0 & c_+\, X & -c_z\, X & c_+\, \Delta X \\
-c_+\, \Delta X & c_z\, \Delta X & c_-\, \Delta X & 0 \\
\end {array} \right)\textrm{,}
\end {equation}
with $c_\pm=c_x \pm ic_y$, $\hat X = \hat X^{(1)} + \hat X^{(2)}$. Here, $\Delta \hat X = \hat X^{(1)} - \hat X^{(2)}$ for the case of two baths and $\Delta \hat X = 0$ for one common bath.

In the following we will, without loss of generality, characterize the results by the angle
$\theta$ between the $\hat \sigma_x$ and $\hat \sigma_z$ component of the coupling
\begin{equation} \label{errcoupling}
\hat \sigma_s \hat X =(\hat \sigma_x\sin \theta + \hat \sigma_z \cos \theta)\hat X\textrm{.}
\end{equation}
This is completely analogous to the bath coupling that is encountered in proposed experimental
qubit realizations, {\em e.g.}, for charge qubits \cite{zorin1}. The bath coupling angle $\theta$ is
defined for $\theta \in [0,\pi/2]$.

Following the lines of \cite {pra} and \cite {governale} the Bloch-Redfield formalism is applied to calculate the
effects of decoherence. The Bloch-Redfield equations and decoherence rates are given analytically.
However, in comparsion to a fully analytic evaluation of the dynamics of the two qubit system \cite{rabenstein},
with this method the time evolution of the reduced density matrix can also be determined numerically for a wide range of system Hamiltonians.

The environment, {\em i.e.}, the bath of harmonic oscillators, is characterized by its spectral
density. The strength of the dissipative effects is given by the dimensionless
parameter $\alpha$.
The bath spectral function is assumed to be linear in frequency up to a cutoff
frequency $\omega_c$. Thus, $J(\omega) = \alpha\hbar\omega/(1+(\omega/\omega_c)^2)$, {\em i.e.}, we
employ an Ohmic spectrum with a Drude cutoff.
The cutoff frequency is chosen two orders of magnitude above the largest frequency
which is typical for a flux qubit system, $\omega_c = 10^{13}$ Hz \cite{thesiscaspar}.

We choose a rather large coupling strength to the environment of $\alpha = 10^{-3}$, which is still in the weak
coupling regime, to be able to observe pronounced decoherence effects.

The Bloch-Redfield equations describe the evolution of the density matrix in the
eigenbasis of the unperturbed Hamiltonian \cite{blum,weiss}
\begin {equation}
\dot{\rho}_{nm} = - i\omega_{nm}\rho_{nm} - \sum_{k\ell}R_{nmk\ell}\rho_{k\ell}\textrm{,}
\end {equation}
where the Redfield tensor $R_{nmk\ell}$ is given by
\begin {equation}
R_{nmk\ell} = \delta_{\ell m}\sum_r\Gamma^{(+)}_{nrrk} + \delta_{nk}\sum_r\Gamma^{(-)}_{\ell rrm} - \Gamma^{(-)}_{\ell mnk} - \Gamma^{(+)}_{\ell mnk}\textrm{,}
\end {equation}
and the rates $\Gamma$ are given by the Golden Rule expressions
\begin {equation}
\Gamma^{(+)}_{\ell mnk} = \hbar^{-2} \int_0^\infty dt \ e^{-i\omega_{nk}t} \langle H_{I,\ell m}(t)H_{I,nk}(0)\rangle_{\beta}
\end {equation}
\begin {equation}
\Gamma^{(-)}_{\ell mnk} = \hbar^{-2} \int_0^\infty dt \ e^{-i\omega_{nk}t} \langle H_{I,\ell m}(0)H_{I,nk}(t)\rangle_{\beta}\textrm{,}
\end {equation}
where $H_{I,\ell m}(t)$ is the matrix element of the bath/system coupling part of the Hamiltonian
in the interaction picture and in the eigenbasis of the decoupled Hamiltonian. Here, $\beta$ indicates averaging
over the degrees of freedom of the thermal bath. In the following $\beta=1/k_B T$, where $T$ is the temperature
of the bath.
Evaluating this, we find according to \cite{pra} for one common bath the rates
\begin{eqnarray}
\Gamma_{\ell m n k}^{(+)} & = & \frac{1}{8\hbar} \Lambda
        J(\omega_{nk}) \left( \coth (\beta  \hbar
        \omega_{nk}/2)- 1 \right) + \frac{i\Lambda}{4\pi \hbar}
        {} \nonumber \\  & &\times {\cal P}
        \int_0^\infty \limits d\omega \ \frac{J(\omega)}
        {\omega^2-\omega_{nk}^2} \left( \coth (\beta \hbar
        \omega/2)\omega_{nk}-\omega \right)\textrm{,}\nonumber\\
\end{eqnarray}
with $\Lambda= \Lambda_{\ell m n k}=\sigma_{s,\ell m}^{(1)} \sigma_{s,nk}^{(1)}
        +\sigma_{s,\ell m}^{(1)} \sigma_{s,nk}^{(2)} +
        \sigma_{s,\ell m}^{(2)} \sigma_{s,nk}^{(1)} +
        \sigma_{s,lm}^{(2)}\sigma_{s,nk}^{(2)}$, and
\begin{eqnarray}
\Gamma_{\ell m n k}^{(-)} & = & \frac{1}{8\hbar} \Lambda
        J(\omega_{\ell m}) \left( \coth (\beta  \hbar
        \omega_{\ell m}/2)+1 \right)  + \frac{i \Lambda}{4\pi
        \hbar} \nonumber\\ & &\times {\cal P}
        \int_0^\infty \limits d\omega \ \frac{J(\omega)}
        {\omega^2-\omega_{\ell m}^2} \left( \coth (\beta \hbar
        \omega/2)\omega_{\ell m} +\omega
        \right)\textrm{.}\nonumber \\
\label{eq:rate_1bath}
\end{eqnarray}
For two distinct baths
\begin{eqnarray}
\Gamma_{\ell m n k}^{(+)} & = & \frac{1}{8\hbar} \left[ \Lambda^1
            J_1(\omega_{nk})+ \Lambda^2 J_2(\omega_{nk})
            \right] {} \nonumber \\ & & \times
            \left( \coth (\beta \hbar \omega_{nk}/2)-1
            \right) {} \nonumber \\ & &{} +
            \frac{i}{4\pi \hbar} \Bigg [ \Lambda^2
            M_2^+(\omega_{nk}) + \Lambda^1 M_1^+(\omega_{nk})
            \Bigg ]\textrm{,}
            \label{gammap_2b}
\end{eqnarray}
with $\Lambda^1 = \Lambda^1_{\ell m n k} = \sigma_{s,\ell m}^{(1)} \sigma_{s,nk}^{(1)}$,
$\Lambda^2 = \Lambda^2_{\ell m n k} =  \sigma_{s,lm}^{(2)} \sigma_{s,nk}^{(2)}$ and
\begin{equation}
M_i^\pm(\Omega) = {\cal P} \int_0^\infty \limits d \omega \
\frac{J_i(\omega)}{\omega^2-\Omega^2} (\coth(\beta \hbar
\omega/2)\Omega \mp \omega)\textrm{,}
\end{equation}
here ${\cal P}$ denotes the principal value. Likewise,
\begin{eqnarray}
\Gamma_{\ell m n k}^{(-)} & = & \frac{1}{8\hbar} \left[ \Lambda^1
            J_1(\omega_{\ell m})+ \Lambda^2
            J_2(\omega_{\ell m}) \right] {}
            \nonumber \\ & & \times \left( \coth (\beta
            \hbar \omega_{\ell m}/2)+1 \right) {}
            \nonumber \\ & &{} + \frac{i}{4\pi \hbar}
            \Bigg [ \Lambda^2 M_2^-(\omega_{lm}) + \Lambda^1
            M_1^-(\omega_{lm}) \Bigg ]\textrm{.} \quad
            \label{gammam_2b} \label{eq:rate_2bath}
\end{eqnarray}
The limit of $\omega_{\ell m}$ tending towards zero can be evaluated separately
\begin {eqnarray}
\Gamma^{(+)}_{\ell mnk} = \Gamma^{(-)}_{\ell mnk} &=& \frac {\alpha}{4\beta\hbar}(\hat \sigma_{s,\ell m}^{(1)}\hat \sigma_{s,nk}^{(1)} + \hat \sigma_{s,\ell m}^{(1)}\hat \sigma_{s,nk}^{(2)} + {} \nonumber \\
 & &{} + \hat \sigma_{s,\ell m}^{(2)}\hat \sigma_{s,nk}^{(1)} + \hat \sigma_{s,\ell m}^{(2)}\hat \sigma_{s,nk}^{(2)})
\end {eqnarray}
for one bath, and
\begin {equation}
\Gamma^{(+)}_{\ell mnk} = \Gamma^{(-)}_{\ell mnk} = \frac {1}{4\beta\hbar}(\alpha_1 \hat \sigma_{s,\ell m}^{(1)}\hat \sigma_{s,nk}^{(1)} + \alpha_2\hat \sigma_{s,\ell m}^{(2)}\hat \sigma_{s,nk}^{(2)})
\end {equation}
for two baths.
All calculations were performed in the same parameter regime as in Ref.~\cite{pra}, thus renormalization effects
of the frequencies are weak and will be neglected.

The ability of a realistic circuit, or in our case a more realistic model of a circuit, to
operate as a quantum gate is characterized by the four gate quality factors introduced
in \cite{pcz}. Those are the fidelity $\mathcal{F}$, purity $\mathcal{P}$, quantum
degree $\mathcal{Q}$ and entanglement capability $\mathcal{C}$. 
The quantum degree and entanglement capability characterize entangling operations. They are 
unique to multi-qubit gates. We will collectively refer to these as non-local gate quality
factors (GQFs) as opposed to fidelity and purity, which are both well defined for an arbitrary
number of qubits, in particular also for a single qubit, and will be referred to as the local gate quality factors.

The fidelity can be evaluated \cite{thorwart} as follows
\begin {equation}
\mathcal{F} = \overline {\bra{\Psi_{\rm in}}\hat{U}^\dagger\hat{\rho}_{\rm out}\hat{U}\ket{\Psi_{\rm in}}}\textrm{.}
\end {equation}
The overline indicates the average over a discrete set of unentangled input states $\ket{\Psi_{\rm in}}$
that can
serve as a basis for all possible input density matrices. The propagator $U$ is the ideal unitary evolution
of the desired gate, and $\hat{\rho}_{\rm out}$ is the density matrix after applying the realistic
gate to $\ket {\Psi_{\rm in}}$. Thus a perfect gate reaches a fidelity of unity and the deviation
from unity indicates the deviation from the ideal process.
The purity $\mathcal{P}$ is indicative of the decoherence effects,
\begin {equation}
\mathcal{P} = \overline {tr(\hat{\rho}_{\rm out}^2)}\textrm{.}
\end {equation}
Again, the overline indicates the input state average. A pure output state leads to $\mathcal{P}$ equal to one,
whereas as the state becomes increasingly
mixed, the square of the weight of the contributions no longer sums up to unity and goes down to
a minimum of one divided by the dimension of the Hilbert space of the system, 1/4 in our case.
\\
Whereas the preceding two factors can be defined for any number of qubits, the following two
are particular to the higher dimensional case
\begin {equation}
\mathcal{Q} = \max_{\rho_{\rm out},\ket{\Psi_{\rm me}}}\bra{\Psi_{\rm me}}\rho_{\rm out}\ket{\Psi_{\rm me}}
\end {equation}
Here the $\rho_{\rm out}$ are the density operators after the gate operation relating to
unentangled input states, whereas the $\ket{\Psi_{\rm me}}$ are the maximally entangled states,
also known as Bell states. Therefore, this measures the ability of the gate to create
quantum entanglement.
\\
Finally, $\mathcal{C}$ is the smallest eigenvalue of the density matrix resulting from
transposing the density matrix of one qubit. As shown in \cite {peres}, the non-negativity of
this smallest eigenvalue is a necessary condition for the separability of the density matrix
into two unentangled systems. After separation, the partially transposed density matrix is a
valid density matrix as well, with non-negative eigenvalues. The negativity of the smallest
eigenvalue thus indicates that the states are not separable and thus non-local.
It approaches -0.5 for the ideal {\sc cnot} gate, the dynamics of entanglement in a
two-qubit system have been studied in Ref. \cite{moore}.
The entanglement capability is closely related to the {\it negativity} $E_N$ of a state
\cite{zyczkowski} which is a non-entropic entanglement monotone \cite{eisert}.

\section{Combination of $\hat \sigma_x$ and $\hat \sigma_z$ errors}
\begin{table}[t]
\begin{center}
\begin {tabular}{c|c|c|c}
{\sc cnot} & local GQFs & nonlocal GQFs & preferred case\\
\hline
\hline
$T \ll T_S$ &close to $\hat \sigma_z$&close to $\hat \sigma_z$& 1 bath\\
\hline
$T \gg T_S$ &close to $\hat \sigma_z$&at $\hat \sigma_x$& 1 bath\\
\multicolumn{4}{c}{}\\
{\sc cphase} & local GQFs & nonlocal GQFs & preferred case\\
\hline
\hline
$T \ll T_S$ &at $\hat \sigma_z$&at $\hat \sigma_z$& --\footnote{There is no
clear tendency observed in this case, see Fig. \ref{GATESlowT}.
Close to pure $\hat \sigma_z$ coupling two baths are preferred,
and close to $\hat \sigma_x$ one bath.}
\\
\hline
$T \gg T_S$ &at $\hat \sigma_z$&at $\hat \sigma_x$& 2 baths\\
\\
\end {tabular}
\end{center}
\caption{Maxima of the gate quality factors for the {\sc cnot} and the {\sc cphase} gate operation. Here $T$ indicates the
temperature and $T_S=E_S/k_B$ is the characteristic temperature scale, which corresponds to the qubit energy scale during the
gate operation. Both the preferred bath configuration and qubit operator coupling to the bath are given.}
\end{table}

\begin{figure*}[t]
\begin{center}
 \includegraphics*[width=1.8\columnwidth]{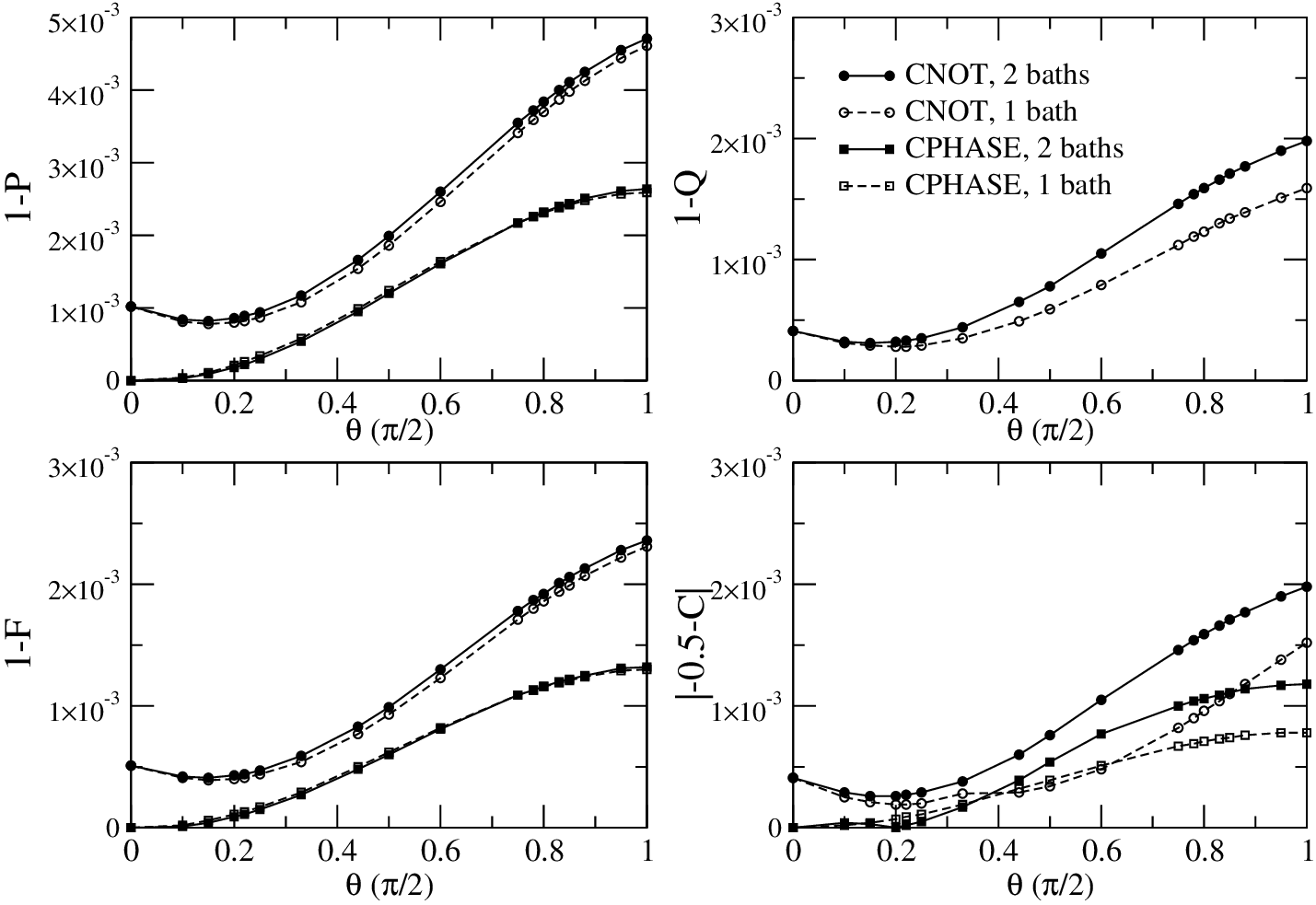}
\end{center}
 \caption{Dependence of the gate quality factors on the bath coupling angle $\theta$ defined in (\ref{errcoupling}) for the
     {\sc cnot} and {\sc cphase} operation at $T \approx 0 \ll T_S$. Here, the behaviour of the gate quality factors for both the single bath and
     two bath case is shown. The characteristic energy scale for the gate operation is $E_S/h=1$ GHz \cite{pra}. The lines are provided as guides to the eye.}
   \label{GATESlowT}
\end{figure*}

\begin{figure*}
 \includegraphics*[width=2.0\columnwidth]{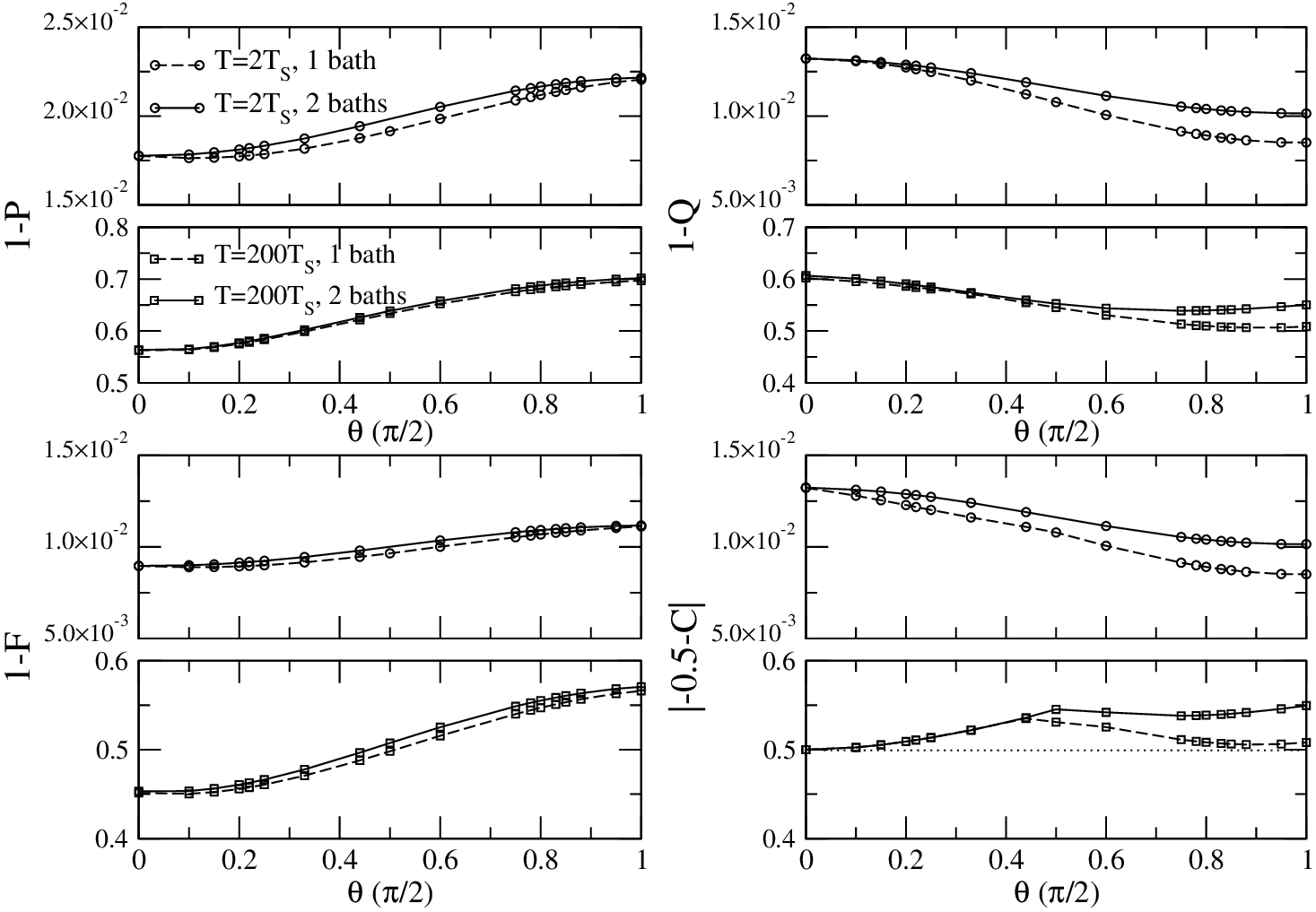}
 \caption{Dependence of the gate quality factors on the bath coupling angle $\theta$ defined in Eqn. (\ref{errcoupling}) for the
    {\sc cnot} operation at large temperatures $T=2T_S$ and $T=200T_S$. The characteristic energy scale for the gate operation is $E_S/h=1$ GHz \cite{pra}. The lines are provided as guides to the eye.}
 \label{CNOT}
\end{figure*}

Now, the spin-boson model with a variable coupling operator to the harmonic oscillator baths
(\ref{errcoupling}) is studied in more detail.

We start with the {\sc cphase} gate which is entangling and forms the core part of the {\sc cnot} operation.
The quantum degree for the {\sc cphase} gate is always smaller than the ideal value
because the {\sc cphase} gate cannot create entangled Bell states. Thus, we did not consider the quantum degree for the
{\sc cphase} gate. The different error coupling configurations achieve the best gate quality factors
for different coupling operators to the environmental baths. The scenarios are summarized in Table 1.

Two qualitatively different temperature regions are found, separated by a smooth crossover. For low temperatures 
$T \ll T_S$, where $E_S/h = (k_B/h) T_S = 1$ GHz is the characteristic energy scale, which
corresponds to the typical qubit energy scale during the quantum gate operation, spontaneous emission processes
dominate.
When $T_S$ is approached, thermal effects become important and for $T \approx T_S$ temperature is the dominating
energy scale as will be discussed in more detail below.

The {\sc cphase} gate, for pure $\hat \sigma_z$ coupling, is protected by symmetry because the gate operation and the coupling to the bath commute.
As was shown previously, all disturbances vanish here in the limit of low temperatures. In this case, spontaneous emission processes are the dominating
decoherence mechanism because absorption and excitation processes are
effectively suppressed due to the bath coupling ({\em i.e.} matrix elements for theses
processes are restricted due to symmetries of the bath coupling operators) and the temperature.
In this case, the low temperature regime can be referred to as the emission
limited regime.

The additional $\hat \sigma_x$ operation in the {\sc cnot} gate Eqn.\,(\ref{cnotgateseq}) during the single
qubit Hadamard operations leads to nonvanishing decoherence rates even
in the low temperature limit. The reason is again the competition of pure dephasing with emission and
absorption processes which show a different dependence on the coupling angle $\theta$.

However, the Hadamard part of the
{\sc cnot} operation is short compared to the overall length of the gate
operation (\ref{cnotgateseq}). Thus, it is found that for low
temperatures the best values for the GQFs are obtained very close to 
pure $\hat \sigma_z$ coupling as depicted in Fig.\,\ref{GATESlowT}. This implies that the overall decoherence
effects are smallest if the bath coupling angle $\theta$ throughout the gate
is adjusted to the distribution of gate operations, which are characterized
by different directions on the Bloch sphere.

For the {\sc cnot} gate better results are observed for the single bath case throughout
the low temperature regime, see Fig.\,\ref{GATESlowT}.
For the single bath configuration, close to pure $\hat \sigma_x$ coupling to the
bath the difference becomes quite significant and for the non local GQFs
actually approaches a factor of two, but reduces again as pure $\hat \sigma_x$
coupling is reached. The {\sc cphase} gate (cf.\,Fig.\,\ref{GATESlowT})
prefers a two bath configuration unless the coupling is $\hat \sigma_x$ dominated.
Again the non-local GQFs are affected most.

For the pure $\hat \sigma_z$ case ($\theta=0$ or correspondingly $c_x=c_y=0$ and $c_z=1$)
little difference between the two bath and the single bath behavior is found in the {\sc cnot} case
and none at all for the {\sc cphase}. It is observed that the single bath configuration is certainly preferred as
soon as there is a significant $\hat \sigma_x$ contribution in the gate operation.
This means that the additional protection from the one dimensional decoherence
free subspace \cite{kempe,pra} involved is mainly beneficial if the commutator of the qubit operator, which couples to the bath
and the qubit Hamiltonian (the Hamiltonian that is needed to perform the individual parts of the gate operation)
has appropriate matrix elements, {\em i.e.}, if there is a significant non-commuting part in the bath coupling
and the gate operations. However, in a $\hat \sigma_z$ dominated case the individual coupling is preferred
as it does not induce any additional indirect couplings between the qubits.
It is natural that the two qubit GQFs should notice this more strongly than
the single qubit GQFs for which the differences never become more than about
one fifth of the individual deviations.
\\
For the high temperature regime, drastically different behavior in the
non-local GQFs is found, see Fig. \ref{CNOT}. Both of them now achieve their best values at
a pure $\hat \sigma_x$ coupling for both gates, the local GQFs achieve their
maximum at a pure $\hat \sigma_z$ value.
The protection that the {\sc cphase} gate enjoyed in the low temperature regime
breaks down here. The high temperature case is essentially scale free,
{\em i.e.}, high temperatures symmetrize the system. In this case the system
eigenbasis is given by the qubit operator which couples to the bath.
This can be nicely shown when considering the single qubit dephasing rates within the spin-boson
model \cite{makhlin1},
\begin{equation}
\Gamma_\varphi = \frac{1}{T_2} = \frac{\epsilon^2}{2E^2}S(0)+
\frac{\Delta^2}{2E^2} S(E)\textrm{.}
\end{equation}
This becomes $\Gamma_\varphi \approx 2 \pi \alpha k_B T/\hbar$ for $T \gg E$ and does not depend on the ratio $\Delta/\epsilon$.

Thermalization is determined by the off-diagonal bath couplings in the basis of the corresponding system Hamiltonian which is required
for a certain gate operation. It will be strongly dominated by the non-commuting contributions, {\em i.e.}, the $\hat \sigma_x$ bath coupling
for the CPHASE part of the CNOT gate. The single-qubit Hadamard part of the CNOT gate will be additionally also affected by the $\hat \sigma_z$ bath coupling.
The two-qubit operation (CPHASE), or in other words the non-local part, of the CNOT gate is of the
$\hat \sigma_z^{(1)} \otimes \hat \sigma_z^{(2)}$-type and the single-qubit Hadamard gates contain both
$\hat \sigma_x$ and $\hat \sigma_z$ contributions. Thus, during the
gate operation the
thermalization is dominated strongly by the $\hat \sigma_x$ part of the
bath coupling for the non-local and
by the $\hat \sigma_z$ part for the local GQFs, implying that for thermal fluctuations the 
$\hat \sigma_x$-type couplings are more important in inducing inter-qubit
transitions, while $\hat \sigma_z$ primarily affects the single qubit gate quantifiers.
What implementation to choose for a gate here, becomes a question of what gate quantifiers are desired to be optimized.
The differences between the one bath and two bath scenario are now small.

For pure $\hat \sigma_z$ coupling of the qubit to the bath, a peculiar effect is observed.
In this case, the minimal eigenvalue of the partially transposed density matrix that is the
entanglement capability stays negative even for $T \gg T_s$ (Fig. \ref{CNOT}). 
The negativity of the eigenvalue of the partially transposed density matrix is not just a necessary
but also a sufficient criterium for the nonseparability of the system in our case \cite{peres}.
Thus, no matter what the temperature or strength of the dissipative effects in our system during the
{\sc cnot} gate operation, entanglement will never be eliminated completely. 
This can be explained by fast thermalization into a protected entangled state.
Furthermore, this effect carries over well into the regime where both $\hat \sigma_x$ and $\hat \sigma_z$ noise are present. 

Overall, the temperature (and the coupling strength) has the largest influence on the GQFs.
At $T \ll T_s$ ($T \approx 10^{-3}$ K), we observe deviations of the GQFs from the ideal value which are less than $10^{-3}$.
At $T \gg T_s$ ($T \approx 1$ K) the deviations are $10^{-1}$ and
quickly increasing further at larger temperatures. The different coupling operators to the bath are the next strongest effect.
Rotating the coupling operator from $\hat \sigma_z$ to $\hat \sigma_x$ causes, in the worst cases, three to four times stronger deviations from the ideal value than the $\hat \sigma_z$ noise. Finally, the change due to different types of bath couplings is generally small compared to the differences between $\hat \sigma_x$ and $\hat \sigma_z$ type coupling. This also suggests that we do not need to worry about noise sources with a weaker coupling strength, even if they couple through a less favorable coupling operator.

As an intermediate conclusion, it is found that for the decoherence dominated regime the {\sc cphase} operation reached the optimum value of the GQFs for a pure $\hat \sigma_z$-coupling to the bath. In the case of the {\sc cnot} operation, the minimum is located slightly shifted to the
$\hat \sigma_x$ component because of the mixture of $\hat \sigma_x$ parts during the
Hadamard operations, compare with Fig.\,\ref{GATESlowT}.

For the {\sc cnot} operation, the optimum values of the four gate quantifiers are
encountered at different bath couplings, which are characterized by the mixing angle Eqn.~(\ref{errcoupling}), especially for large temperatures.

Thus, the differences between the case of one common bath and two baths are much less important than the
symmetries between the gate operation and the bath operators. In particular, the difference
between the case of one or two baths disappears for pure $\hat \sigma_z$ coupling to the bath. Here, decoherence due to flux noise or charge noise in coupled
superconducting flux or charge qubits was explored. Decoherence
due to $1/f$ noise, caused by background charges or bistable fluctuators, was
not treated.

\section{Non-identical Qubits}

\begin{figure}[t]
 \begin{center}
  \includegraphics*[width=8.6cm]{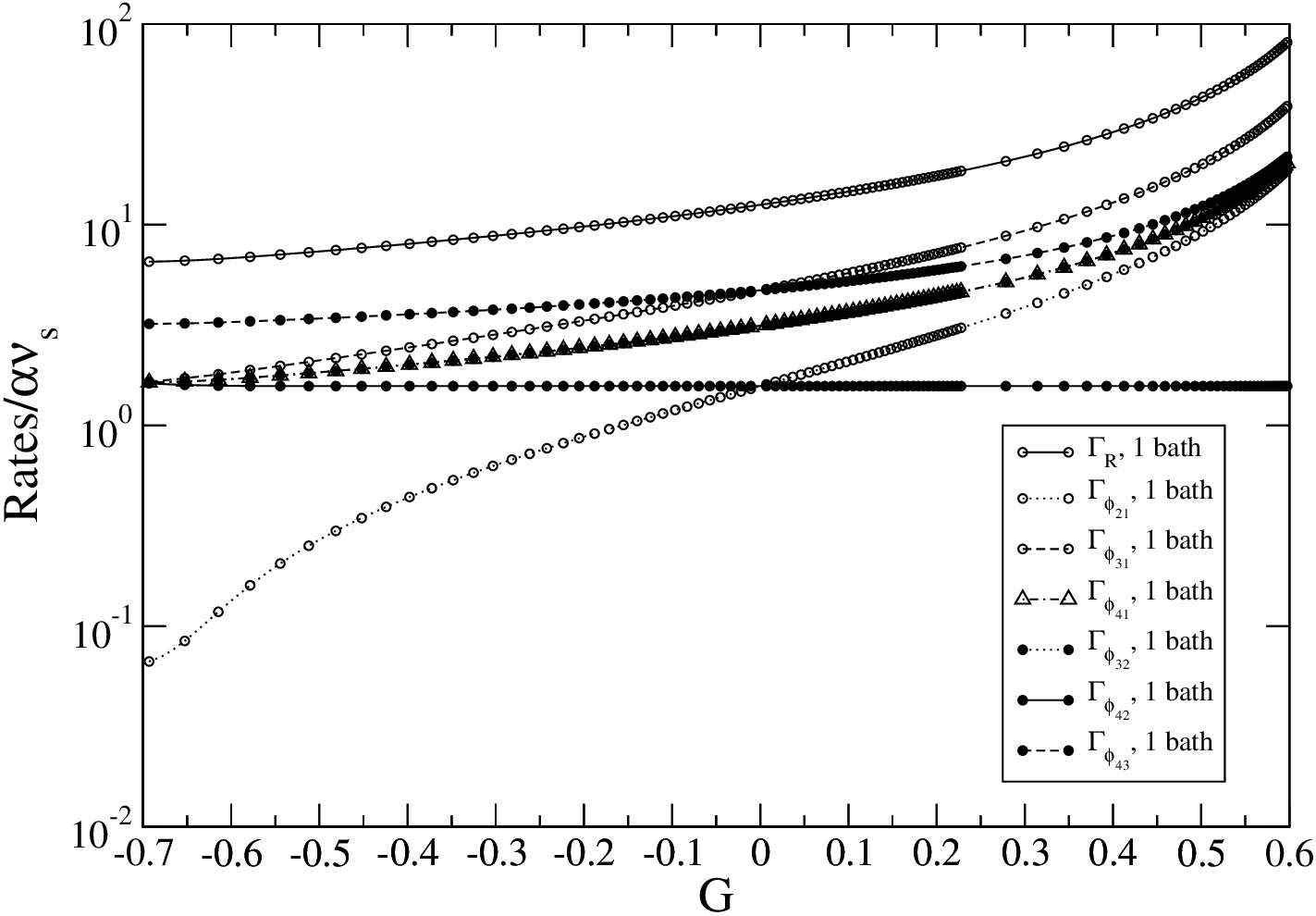}
 \end{center}
\caption{Dependence of the decoherence rates on the qubit asymmetry. Here, we set $K=0$, $\epsilon_i=0$, $\Delta_1=E_S$, and vary $\Delta_2$.
The single and two bath cases behave identically, thus only the single bath case is shown. The strength of the decoherence effects is set to $\alpha=10^{-3}$ and $T \approx 0.5 T_S$.
We set the bath coupling angles (\ref{asbc}) to $\theta_1=0$ and $\theta_2=0$. The decoherence rates are scaled by $\alpha \nu_S$, with $\nu_S=E_S/h$. The lines are provided as guides to the eye.}
\label{Asymmetry}
\end{figure}

\begin{figure}[t]
\begin{center}
 \includegraphics*[width=8.6cm]{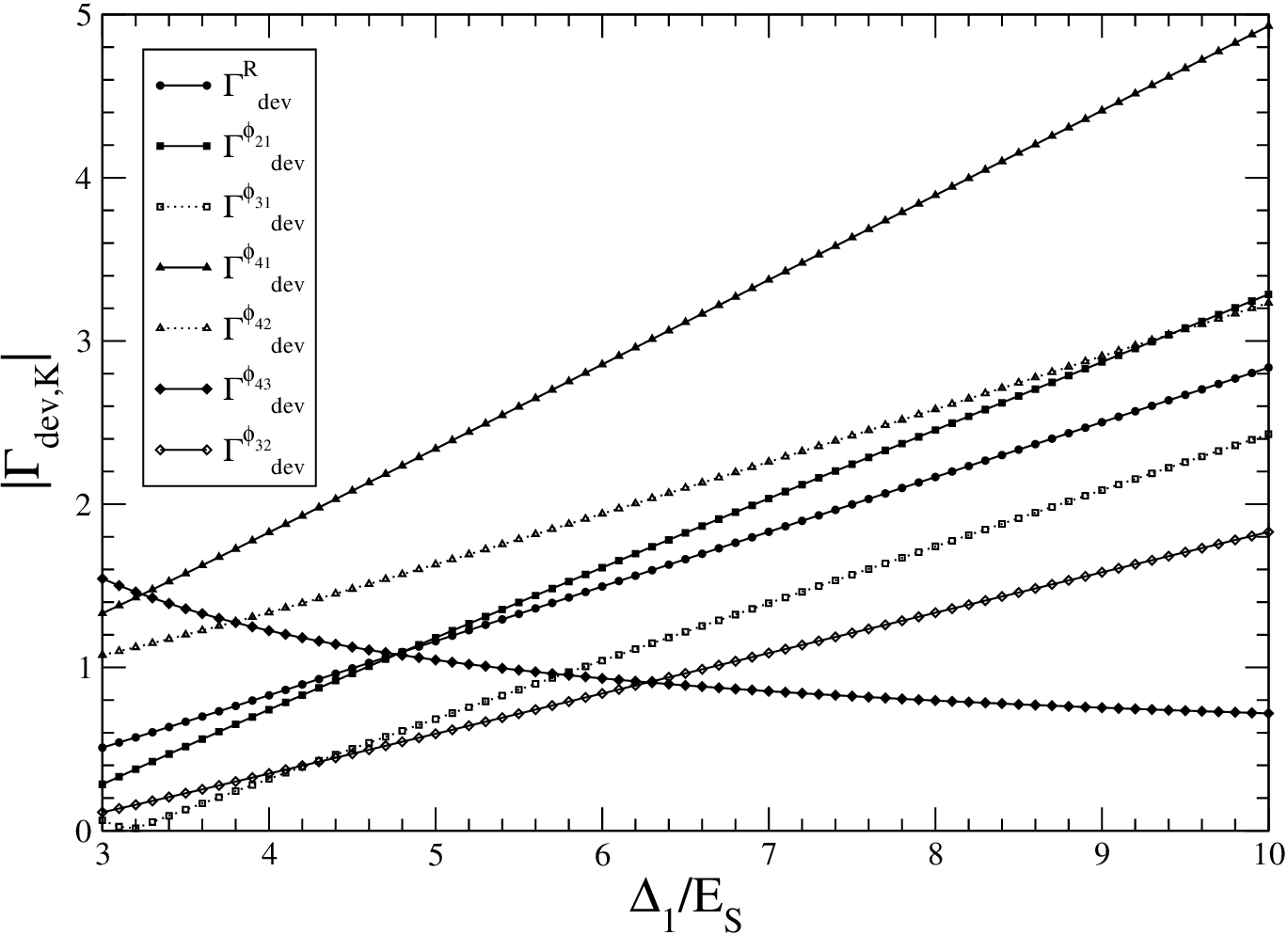}
\end{center}
 \caption{Dependence of the decoherence rates on the qubit asymmetry at $T\approx 0.5T_S$. Here, the case of one common
bath is investigated. The tunnel matrix element of the second qubit and the inter-qubit coupling are set to $\Delta_2=K=E_S$ and $\Delta_1$ is varied. 
For comparison with experiments, large asymmetry in the tunnel matrix elements of the individual qubits is investigated.
The bath coupling angles are set to $\theta_1=0$ and $\theta_2=0$. The strength of
the dissipative effects is $\alpha=10^{-3}$. The lines are provided as guides to the eye.}
   \label{asym_1bath}
\end{figure}

\begin{figure*}[t]
\begin{center}
 \includegraphics*[width=2.0\columnwidth]{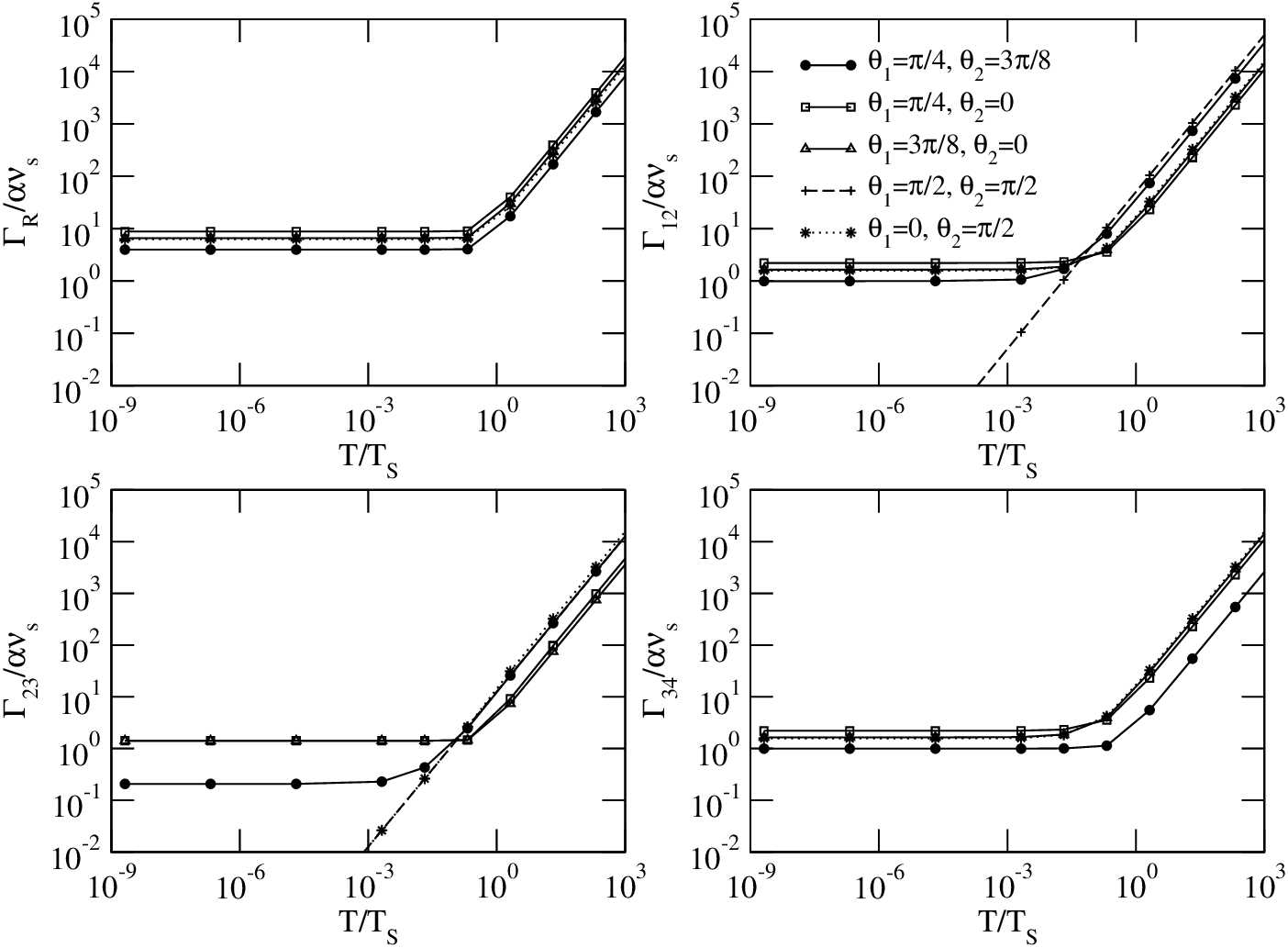}
\end{center}
   \caption{Temperature dependence of selected decoherence rates for $K=0$, $\epsilon_i=0$, $\Delta_1=E_S$, $\Delta_2=0.9 \Delta_1$, for the single bath case. Here, temperature and the bath
coupling angles are varied. The strength of the dissipative effects is set to $\alpha=10^{-3}$. The decoherence rates are scaled by $\alpha \nu_S$, where
$\nu_S=E_S/h$. The lines are provided as
guides to the eye.}
   \label{TdepRates}
\end{figure*}

Now, we do not restrict the analysis to the case of a uniform error coupling (\ref{errcoupling}) anymore. In general, both qubits can
couple to the baths differently
\begin{equation} \label{asbc}
\hat \sigma^{(i)}_s = \hat \sigma_x^{(i)} \sin \theta_i + \hat \sigma_z^{(i)} \cos \theta_i\textrm{,}
\end{equation}
where $i=1,2$ denotes one of the qubits.
For the numerical calculations, the qubits are set to the degeneracy point $K=\epsilon_i=0$ and $\Delta_1$ is set to $E_S$, {\em i.e.}, effectively this model describes
a system of two uncoupled qubits. Here, $\Delta_2$ and thus also the asymmetry $G=(\Delta_1-\Delta_2)/(\Delta_1+\Delta_2)$ is varied.

Experimentally the spread of the qubit parameters due to fabricational imprecision is very important both
because quantum algorithms (without further modification) require a certain level of precision and the decoherence
effects in the system of qubits have to be sufficiently small \cite{aharonov}.

Therefore, it is of central importance to investigate also the effects of the parameter spread in non-identical qubits on
the behaviour of the decoherence rates. Superconducting qubits are preferably operated at the degeneracy point where decoherence
effects are suppressed for superconducting charge and flux qubits. However, the tunnel matrix elements for superconducting qubits
can differ significantly,  on the order of several percent \cite{cnotscb,paauw}. Thus, the dependence of the decoherence rates,
{\em i.e.}, dephasing rates and the relaxation rate, close to the degeneracy point on the qubit asymmetry is an important property.

Figure~\ref{Asymmetry} depicts the dependence of the decoherence rates on the qubit asymmetry $G$ when the individual qubits are
operated close to the degeneracy point. Temperatures typical for experimental situations are chosen for this analysis, {\em i.e.},
$T \approx 0.5 T_S$.

We observe that for pure $\hat \sigma_x$ coupling
to the bath ($\theta_1=\pi/2$ and $\theta_2=\pi/2$), the asymmetry of the qubits is irrelevant because the coupling to the bath and the system Hamiltonian commute
(the indices of qubit one and qubit two could be exchanged without changing the system). For mixed bath coupling
of the $\hat \sigma_x$ type for one qubit ($\theta_1=\pi/2$) and the $\hat \sigma_z$ ($\theta_2=0$) type for the other qubit, still the decoherence rates do not vary for different
asymmetry. The reason for this behaviour is that the $\hat \sigma_x$ bath coupling of the second qubit {\it always} commutes with
the qubit Hamiltonian, {\em i.e.}, only flipless dephasing processes contribute to the decoherence rates of the second qubit. When we vary the
asymmetry, essentially $\Delta_2$ is varied ($\Delta_1=E_S$ is kept constant), which leads to a different contribution of the second qubit
to the overall decoherence. However, these corrections are small
compared to the full decoherence rates ({\em i.e.}, not only flipless dephasing processes) that contribute in the
case of qubit one where $[H_{\rm SB},H_{\rm sys}] \neq 0$ and $\Delta_1$ stays constant.

Finally, Fig.\,\ref{Asymmetry} shows the case of fully perpendicular system Hamiltonian and bath coupling. Here, the decoherence rates increase steeply
for increasing asymmetry. Note here that due to the definition of $G$ and $\Delta_1$, the two cases $G=-0.8 \rightarrow \Delta_2=9 E_S$
and $G=0.8 \rightarrow \Delta_2=1.1 E_S$ are vastly different.

From the dependence of the decoherence rates on the asymmetry $G$ of the two qubits
at the degeneracy point it is possible to estimate the maximum tolerable asymmetry for a given constraint
on the deviation of the decoherence rates from their value for perfectly symmetric qubits.
We define the deviation of the decoherence, {\em i.e.}, the relaxation or dephasing rates from
their values at the degeneracy point as
\begin{equation} \label{gammadev}
\Gamma_{\rm dev}^{R,\varphi_{ij}} = 1 - \frac{\Gamma_{R,\varphi_{ij}}(\Delta_1 \neq \Delta_2,K=\epsilon_i=0)}{\Gamma_{R,\varphi_{ij}}(\Delta_1=\Delta_2,K=\epsilon_i=0)}\textrm{.}
\end{equation}
We find in the case of a bath coupling parallel to the Hamiltonian that for $\Gamma_{\rm dev}^R$ to be smaller than 1\%, it is required that
$0.5 \Delta_2 < \Delta_1 < 1.5 \Delta_2$, {\em i.e.},
the parameter spread of the two qubits could be remarkably large ($\approx$ 50\%)
without considerably affecting the relaxation rates. However, detailed analysis
shows that for the $\Gamma_{\rm dev}^{\varphi_{ij}}$, {\em i.e.} the deviations of the dephasing rates, the
increase happens much earlier. {\em E.g.}, in the case of $\Gamma_{\rm dev}^{\varphi_{32}}$ already for a deviation between
$\Delta_1$ and $\Delta_2$ of less than approximately 10\%. Moreover, there is a large spread among
the dephasing rates, which are sensitive to the qubit asymmetries.
Note that both the single bath and the two bath case behave identically for the relaxation rate.
Differences between the two cases only occur for the dephasing rates.
The angles of the bath coupling where the minimum dephasing rates are encountered are
different for the different dephasing rates.

Typical experimental values for charge \cite{cnotscb} and flux qubit \cite{paauw} designs indicate that the
parameter spread in the tunnel matrix amplitudes can be quite large, in the case of the charge qubit it is a
factor of $\Delta_2/\Delta_1 \approx 0.91$ and for the flux qubit $\Delta_2/\Delta_1 \approx 4.22$. This difference of asymmetries is due to the fact,
that fabrication parameters such as $E_c$ and $E_J$ enter the {\em exponent}
of the tunnel splitting in the flux qubit case \cite{orlando}.
These experimental values emphasize that it is important to study the evolution of the decoherence effects
for non-identical qubit parameters. Moreover, important information about the noise sources coupling to
the qubit can be identified. From comparison of the decoherence rates for different qubit samples, 
which possess different asymmetries between the tunnel amplitudes, it is thus possible to identify
the predominant bath coupling angle. In most qubit designs the bath coupling angle is then uniquely related
to a certain noise source, {\em e.g.}, flux noise in the case of flux qubits \cite{epj03}.

Figure~\ref{asym_1bath} depicts the experimentally important \cite{paauw} behaviour of the decoherence rates when $\Delta_2$ and $K$ are
fixed to $E_S$  and $\Delta_1> \Delta_2$ is changed. Here, different from the definition in Eqn. (\ref{gammadev}), all rates are
normalized by $\Gamma_{R,\varphi_{ij}}(\Delta_1=\Delta_2=K,\epsilon_i=0)$, which will be labelled by $\Gamma_{\rm dev,K}$.
In this case the two qubits are permanently coupled, and embedded into one common environmental bath.
The decoherence rates begin to increase linearly when $\Delta_1$ is larger than $\Delta_2$.

Figure~\ref{TdepRates} illustrates the temperature dependence of the decoherence rates for
the case of one common bath. The values of the decoherence rates for the two bath case differ
insignificantly from the single bath case.
We observe that the spread of the magnitude of the different decoherence rates increases at intermediate mixing angles. As expected, the magnitude
of the decoherence rates is maximum in the case where the system Hamiltonian and the coupling to the bath are perpendicular to each other. For the
opposite case, where the system Hamiltonian and the coupling to the bath commute, the
decoherence rates vanish for decreasing temperature, {\em i.e.}, only flipless dephasing processes contribute to the overall decoherence.
Note that in the case where the system Hamiltonian and the coupling to the bath commute ($\theta_1=\pi/2$ and $\theta_2=\pi/2$) the relaxation rate
vanishes.

It is found that the dephasing rates depend strongly on the qubit asymmetry.
Nevertheless the parameter spread of the qubit energies can be quite large (around 10\%) without
affecting the decoherence properties considerably. However, for very large asymmetries and
a bath coupling, which is perpendicular to the system Hamiltonian, the decoherence rates
increase exponentially with asymmetry. 

\section{Conclusions}
A system of two pseudospins coupled by an Ising type zz-interaction, which models {\em e.g.} superconducting
charge or flux qubits, was investigated.
It was shown that the for the system of two pseudospins the optimum gate performance of different gate operations is closely
related to their composition of elementary gates and the coupling to the bath.
When considering the gate quality factors, the temperature and special symmetries of
the system-bath coupling have the largest influence on the decoherence properties, whereas the single
or two bath scenarios make only little difference.

For the {\sc cphase} operation at low temperatures, the optimum gate quality factors are
at pure $\hat \sigma_z$ system-bath coupling due to the fact that only in this case
all individual Hamiltonians necessary for performing the gate operations and the
system bath coupling commute. Similarily, the {\sc cnot} gate operation approaches the
best gate quality factors close to $\hat \sigma_z$ system-bath coupling with a slight
$\hat \sigma_x$ admixture due to the Hadamard operations.

For very large temperatures, the temperature effectively symmetrizes the system and
thus entanglement is always preserved during a {\sc cnot} gate operation independently of the
system-bath coupling.

Furthermore, it is found that the parameter spread of the tunnel matrix elements of the
qubits, when operated close to the degeneracy point can be quite large (approximately 10\%) for
a bath coupling which commutes with the system Hamiltonian
without affecting the decoherence properties considerably.
In this case the differences in the decoherence rates stay below 1\%.

This work was supported in part by the NSA and ARDA under ARO contract
number P-43385-PH-QC and by DFG through
SFB 631. We thank A.B.~Zorin, W.D.~Oliver, A.~Marx, L.C.L.~Hollenberg, S.~Kohler, U.~Hartmann, and M.~Mariantoni for useful discussions.

\end{document}